\documentclass[12pt, letterpaper]{article}
\usepackage[top=1in, bottom=1.5in, left=1in, right=1in]{geometry}
\usepackage[utf8]{inputenc}
\usepackage{booktabs}
\usepackage{hyperref}
\usepackage{graphicx}
\title{Blazars and Fast Radio Bursts with LSST}
\author{C.M. Raiteri$^1$\thanks{claudia.raiteri@inaf.it},  M.I. Carnerero$^1$, B. Balmaverde$^1$, F. D'Ammando$^2$, C. Righi$^3$, \and A. Possenti$^4$, E. Pian$^5$, A. Capetti$^1$, M. Villata$^1$, M. Giroletti$^2$, P. Romano$^3$, \and A. Stamerra$^6$, F. Tavecchio$^3$, S. Vercellone$^3$
\and with the support of the LSST Transient and Variable Stars Collaboration
\and {\small {\it $^1$ INAF-Osservatorio Astrofisico di Torino, Italy}}
\and {\small {\it $^2$ INAF-Istituto di Radioastronomia di Bologna, Italy}}
\and {\small {\it $^3$ INAF-Osservatorio Astronomico di Brera, Italy}}
\and {\small {\it $^4$ INAF-Osservatorio Astronomico di Cagliari, Italy}}
\and {\small {\it $^5$ INAF-Osservatorio di Astrofisica e Scienza dello Spazio di Bologna, Italy}}
\and {\small {\it $^6$ INAF-Osservatorio Astronomico di Roma, Italy}}
}
\date{November 30, 2018}

\begin{document}

\maketitle

\begin{abstract}
The aim of this white paper is to discuss the observing strategies for the LSST Wide-Fast-Deep that would improve the study of blazars (emission variability, census, environment) and Fast Radio Bursts (FRBs). For blazars, these include the adoption of: i) a reference filter to allow reconstruction of a well-sampled light curve not affected by colour changes effects; ii) two snapshots/visit with different exposure times to avoid saturation during flaring states; iii) a rolling cadence to get better-sampled light curves at least in some time intervals.  
We also address the potential importance of Target of Opportunity (ToO) observations of blazar neutrino sources, and the advantages of a Minisurvey with a star trail cadence (see white paper by David Thomas et al.) for both the blazar science and the detection of possible very fast optical counterparts of FRBs. 
\end{abstract}

\section{White Paper Information}
\begin{enumerate} 
\item {\bf Science Category:} Exploring the transient optical sky
\item {\bf Survey Type Category:} Wide-Fast-Deep and Minisurveys 
\item {\bf Observing Strategy Category:} Blazar and FRBs science is best described as a specific observing strategy to enable specific time domain science, that is relatively agnostic to where the telescope is pointed. 
\end{enumerate}

\clearpage

\section{Scientific Motivation}

Blazars are radio-loud active galactic nuclei (AGN) with one jet pointing towards us, so that the jet emission is relativistically beamed. The main effects are an enhancement of the flux, a contraction of the variability time scales, and a blue-shift of the frequencies. 

Blazar emission shows unpredictable variability over all the electromagnetic spectrum, from the radio band to gamma rays, with time scales ranging from minutes to years. Polarimetric observations in radio and optical bands show that the flux is highly polarised, and that both polarization degree and angle are variable too.

Blazars come in different flavours.
Depending on the strength of emission lines in their optical spectra, we classically distinguish between flat-spectrum radio quasars (FSRQs), with strong broad emission lines, and BL Lac objects (BL Lacs), with (almost) featureless continua, the difference being probably due to different regimes of the accretion flow. 
For both types the spectral energy distribution (SED) is dominated by the non-thermal radiation from the jet and displays two bumps in the usual $\log \nu F_\nu$ versus $\log \nu$ plot. The low-energy bump is due to synchrotron radiation, while the high-energy bump likely to inverse-Compton emission and/or to hadronic processes.
Based on the position of the SED peaks we further divide BL Lacs into low-energy-peaked BL Lacs (LBL) and high-energy-peaked BL Lacs (HBL). These latter emit photons up to TeV frequencies.

Multiwavelength observations can greatly help the study of blazar variability and identification and characterization of new blazars. The latter task is facilitated by spectroscopic observations, while polarimetric data can be very useful too, as they give information on the behaviour of the magnetic field.
The future observations of the Square Kilometer Array (SKA) in the radio band and those of the Cherenkov Telescope Array (CTA) at TeV energies will in particular give a strong support to LSST for blazar studies.

Known blazars are relatively rare objects. 
The BZCAT5 catalogue\footnote{https://www.asdc.asi.it/bzcat/} includes 3561 objects, up to a known redshift of 6.8 and maximum magnitude $R\sim 24$, which have been selected in different ways. The 2WHSP catalogue\footnote{https://www.ssdc.asi.it/2whsp/} contains 1691 sources, all HBL up to a known redshift of 0.8, 553 of which already included in BZCAT5. Therefore, we are dealing with about 4700 objects.
However, due to the beaming effect on the jet emission, {\bf blazars are the best objects to study what is happening in the inner regions of extragalactic jets, close to the supermassive black hole}. {\bf LSST will allow us to identify many more faint objects, most of them located at very high redshifts} (there are only 80 known blazars with $z>3$). It will then be possible to {\bf obtain constraints on the growth of supermassive black holes in the early Universe} (Ghisellini et al. 2013).

We have a biased knowledge of the blazar emission behaviour, since observing campaigns usually target the brightest and most variable sources, specially during outburst phases (e.g.\ Raiteri et al. 2017). 
{\bf LSST will provide a long-term, multiband monitoring of about 2000 already known blazars and of a number of newly discovered ones to study their variability properties in an unbiased way.}
In particular, we will be able to {\bf verify the presence of (quasi)periodic behaviour} that, depending on the time scales, can {\bf reveal jet precession or even the presence of a binary black hole system} (BBHS). A BBHS is indeed expected if the giant elliptical galaxies hosting blazars are produced through a {\bf galaxy merging process}. 
Up to now there have been several clues of (quasi)periodicity of blazar light curves in various bands, the most impressive being the centenary optical light curve of OJ 287, which is strongly suspected to host a BBHS (e.g. Dey et al. 2018), but the statistical relevance of the phenomenon is still to be ascertained.

{\bf Blazar jets are powerful cosmic accelerators}. Indeed, blazars represent the most abundant extragalactic population at gamma rays, as confirmed by the {\it Fermi} satellite, and potential high-energy neutrino and cosmic rays emitters.
There have been a number of works attempting to connect neutrino events detected by IceCube with blazars, in particular HBL, and recently we had the first robust association of a high-energy neutrino, IceCube-170922A, with a flaring blazar, TXS 0506+056 at redshift $z=0.34$ (The IceCube Collaboration et al. 2018).
This confirmed that blazars may indeed be the {\bf birthplace of the most energetic particles in the Universe}, allowing us to explore the most energetic Universe to distances from where photons cannot reach us because they are absorbed.
However, more associations must be found before we can estimate what fraction of the high-energy neutrinos and cosmic rays are due to blazar activity and under which conditions these particles are produced.

We notice that most {\bf unidentified $\gamma$-ray sources} detected at GeV energies by the {\it Fermi} satellite and at TeV energies by Cherenkov telescopes are expected to be blazars. Together with a possible spectroscopic follow-up, {\bf LSST will confirm blazar candidates and help to identify the other objects.} 

Blazar variability studies imply more stringent constraints on sampling rather than on depth. Depth however is fundamental for {\bf environmental studies}. Radio-emitting AGN are often found in rich environments (Kotyla et al. 2016) and the most powerful of them are in {\bf galaxy clusters}, those with the most massive hosts being located in the central regions (Magliocchetti et al. 2018). We would expect that blazars share the same clustering properties of their parent population, the radio galaxies, but this has to be verified with {\bf deep images of blazars and radio galaxies environments that LSST will provide}.

LSST can potentially play a major role also in the identification of {\bf Fast Radio Bursts} (FRBs) counterparts. FRBs are radio transients with millisecond duration of still unknown nature. They were discovered about ten years ago, but there has been just one robust identification up to now: the repeating FRB121102 (Spitler et al. 2016), which  likely comes from a dwarf galaxy at redshift 0.2. Many models have been proposed to explain FRBs, most of them involving neutron stars at high redshifts. Only a few tens of events have been detected, but the real rate is expected to be of several thousands a day. We plan to explore the FRBs position error boxes of radio surveys (e.g. those provided by the SUPERB consortium, Keane et al. 2018, or by additional experiments about to start) on LSST images to identify possible optical counterparts.

\begin{center}
\includegraphics[scale=0.6]{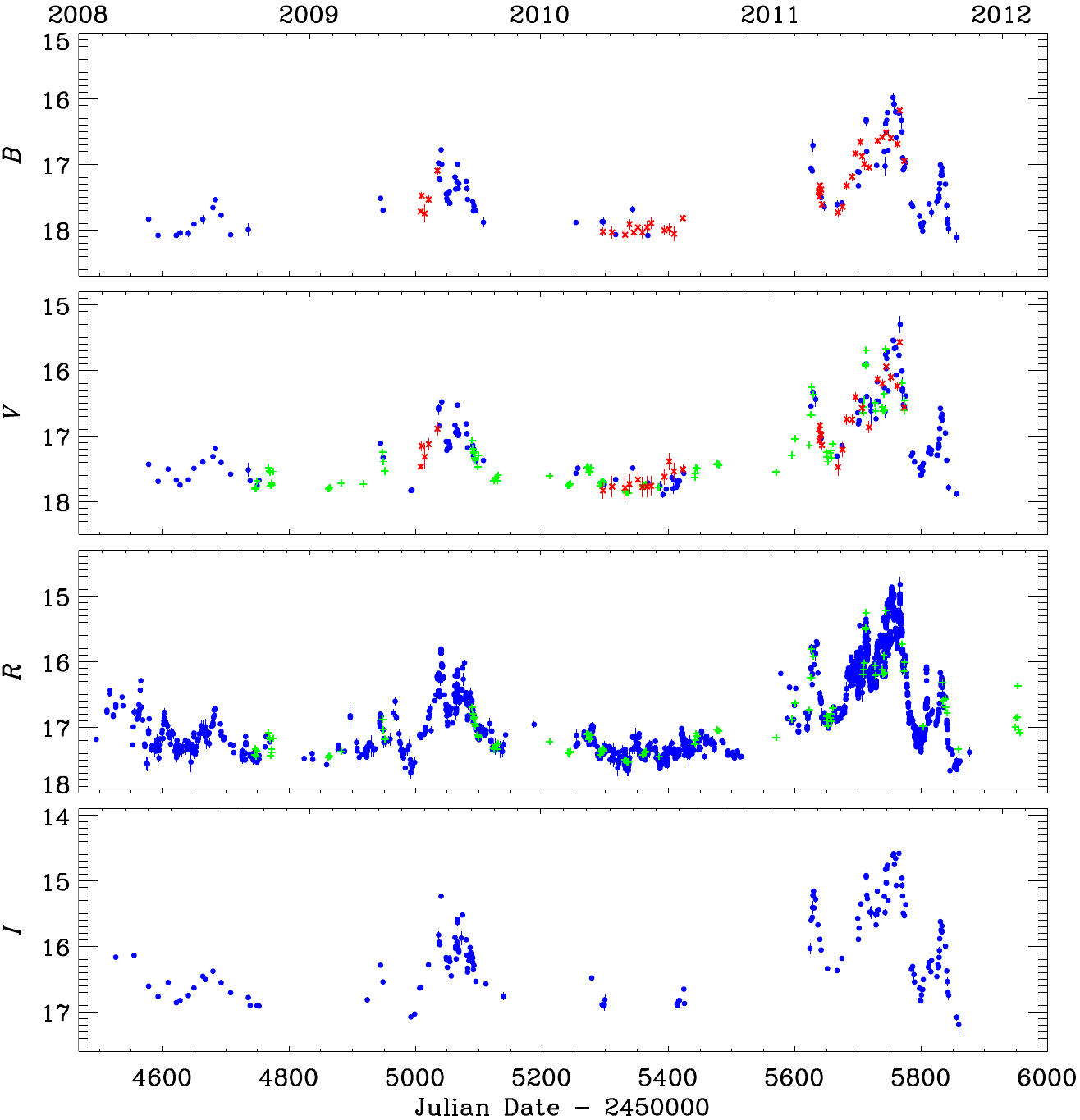}
\includegraphics[scale=0.5]{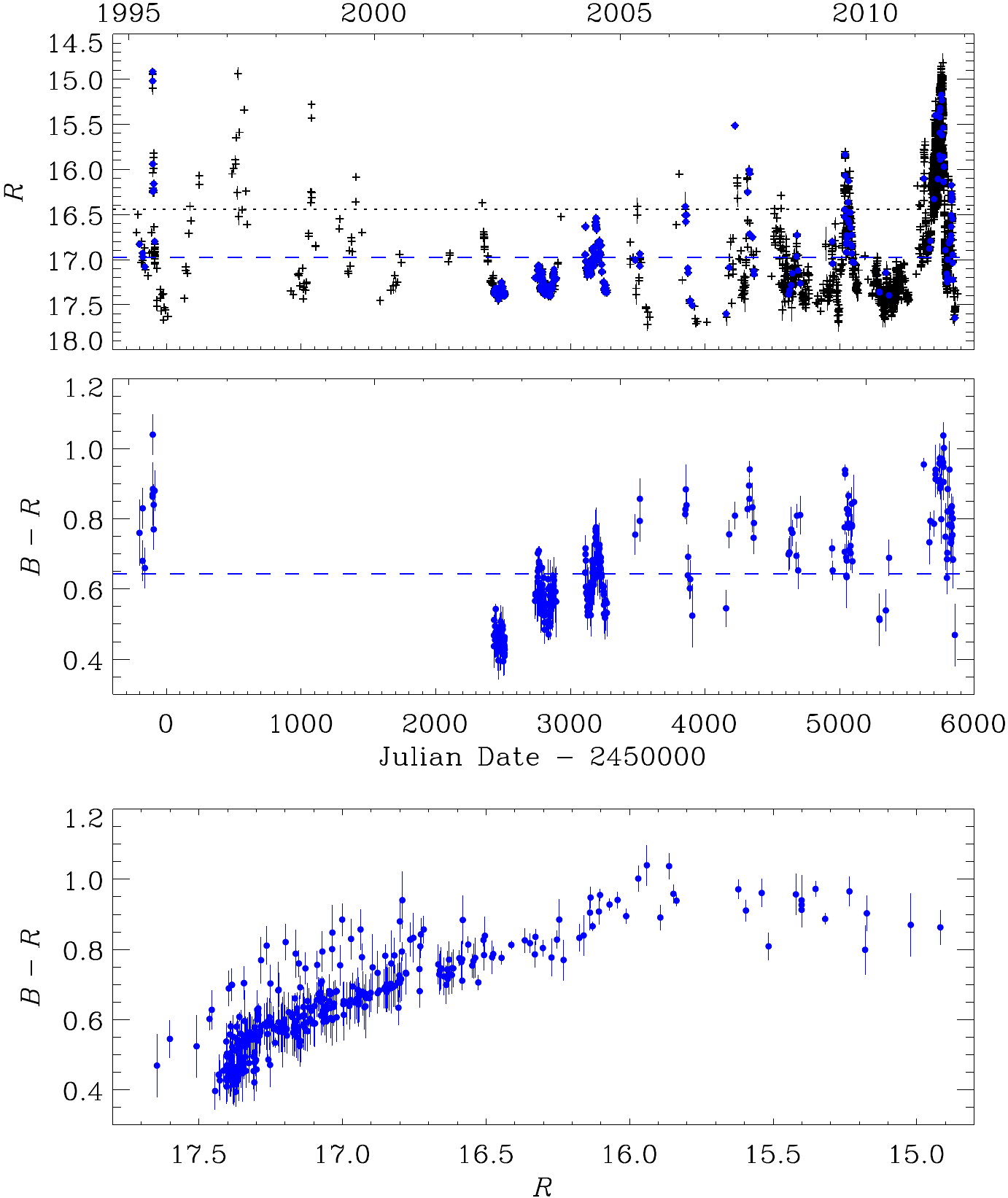}\\
\vspace{0.2cm}
\includegraphics[scale=0.45]{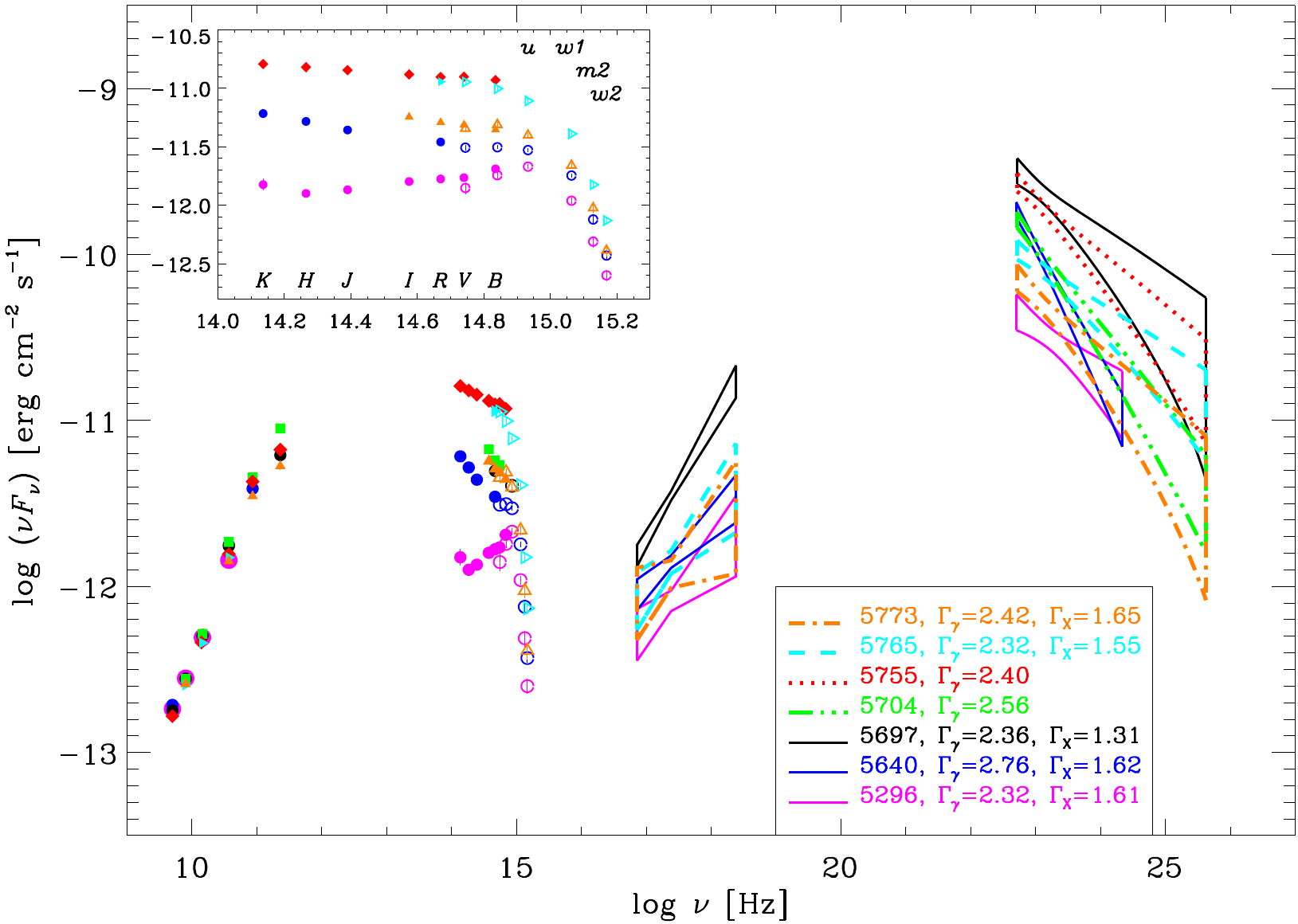}
\includegraphics[scale=0.45]{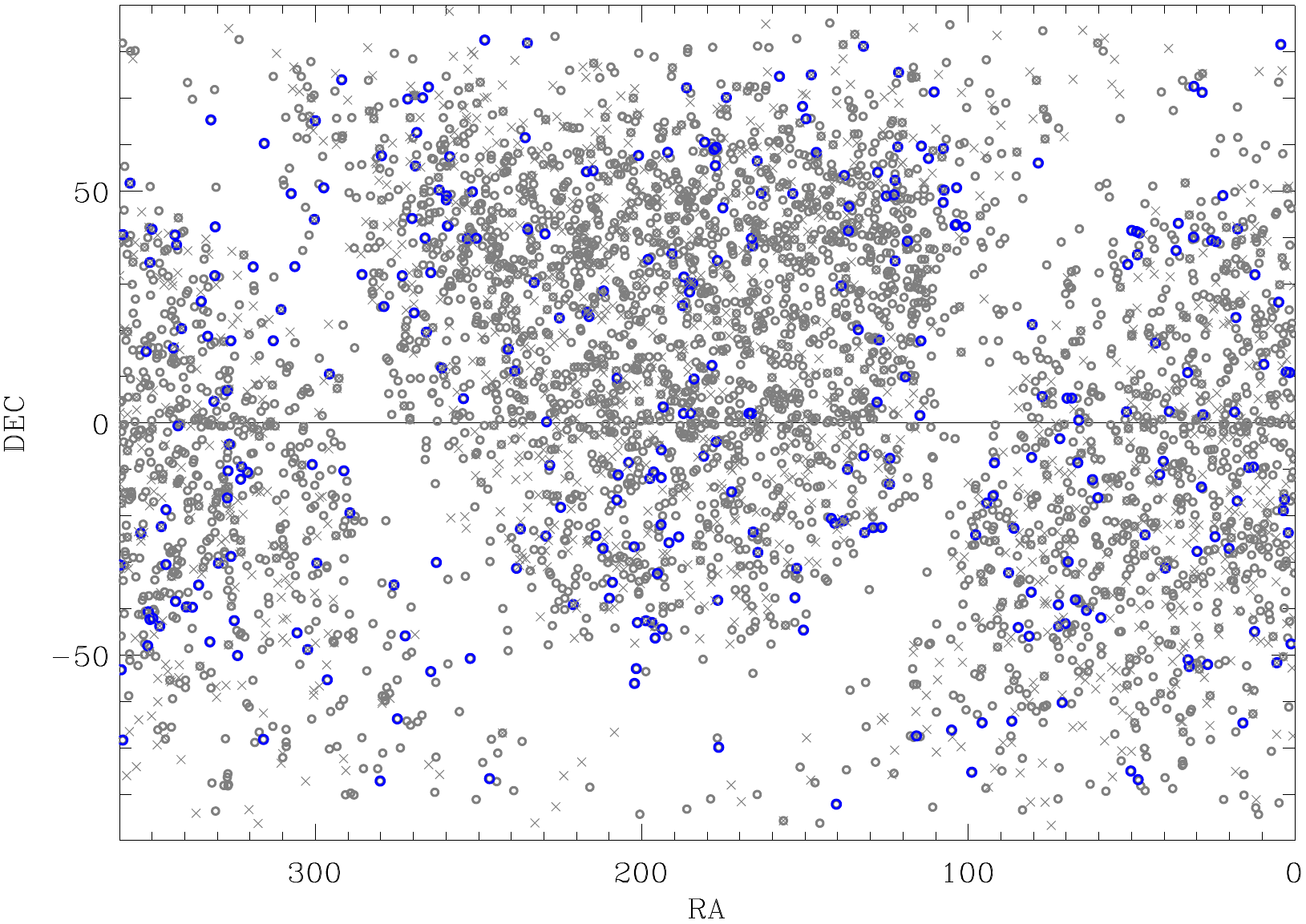}
\end{center}

{\it In the above figure we show an example of the optical and spectral variability of a blazar, the FSRQ 4C 38.41 (1633+382) at redshift $z=1.814$. 
The top left panel displays the optical light curves, the top right panel the colour behaviour, and the bottom left panel the broad-band (radio-to-$\gamma$-rays) SEDs in different epochs, corresponding to different brightness states (from Raiteri et al. 2012). The bottom right panel shows the location of the blazars known (grey symbols) and of those with brightness exceeding the LSST saturation limit in $r$ band for a 15 sec exposure (blue circles).}

\vspace{.6in}

\section{Technical Description}

LSST, with its unique capability to provide a continuous and 10-years long monitoring of the whole Southern Sky with unprecedented depth, will be a formidable tool to shed light on many blazars and FRBs open issues. 
However, details of the LSST observing strategy can make a difference on the results that can be obtained. 
The aim of this white paper is to highlight the best observing choices for the blazar and FRB science cases.

{\bf Wide-Fast-Deep}

In general, a few day sampling is reasonable to follow blazar variability and to potentially identify new objects. 
Exceptions are flaring and outburst phases, when the time scales on which the flux changes usually shorten. 
In principle, Deep Drilling Fields (DDFs) would be more appropriate to analyse blazar variability during very active phases, but the brightness involved together with the blazar paucity prevent us from proposing DDFs for the blazar science. We incidentally notice that among the 4 already selected DDFs, 3 do not contain known blazars, while the fourth, COSMOS, contains only 2 faint objects ($V=19$--21).

One major issue is that blazars undergo spectral changes (see figure in section 2), so it is not possible to build a reliable light curve from observations in different bands. 
We performed a MAF simulation with the TransientAsciiMetric to highlight this. We used the $griz$ (actually $BVRI$) light curves of 4C 38.41 (see figure in section 2) as input for the TransientAsciiMetric and derived the corresponding light curves obtained by LSST when considering the baseline\textunderscore 2018a OpSim Run. Then we adopted average colour indices to reconstruct a composite $r$-band light curve. The result is shown in the figure below, where we see a large mismatch between the data converted into $r$ mag and the true $r$-band values.

\begin{center}
\includegraphics[scale=0.5]{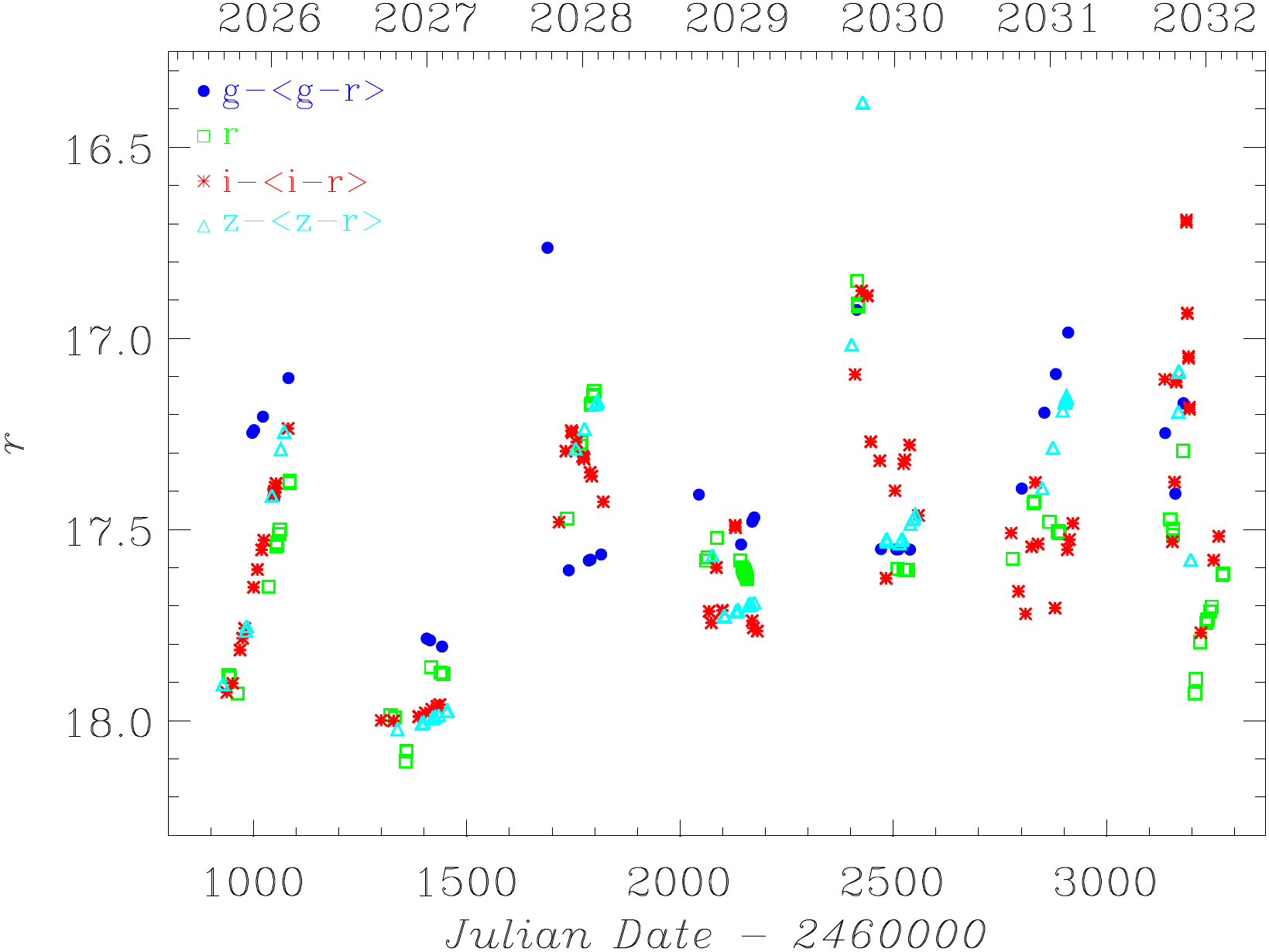}
\end{center}

Therefore, we would need a reference filter, so that we can have a monochromatic light curve sampled every few days. This may be accomplished if {\bf one of the two planned visits/night of each field is always performed in a reference filter} (e.g. the $r$ filter).

However, since this strategy would penalise the detection of transients with variability time scales of about 1 hour, the best strategy would probably be to go for 3 visits/night, one of which in the reference filter.
As for the other filters, since we have both ``red" and ``blue" objects spread over a wide range of redshifts, all filters are important to characterize their properties, 
so we do not give specific prescriptions for favouring some of them with respect to the others. 

Another major problem is {\bf saturation}. The limit for a 15 sec snapshot is $r=16$, which is already a heavy constraint for blazar science. During outbusts, blazars can brighten up to 6 magnitudes (Raiteri et al. 2017), so even a faint object can reach the saturation limit, when it is flaring. Yet, flaring states are the most important ones to understand the jet physics, also because they can be more easily followed by facilities at other wavelengths, in particular at high energies (e.g. CTA, which will detect hundreds of blazars in the TeV regime). 
Visits with single 30 sec snapshots would have an even worse impact.
Therefore, we propose to {\bf diversify as much as possible the two exposures of a single visit}. A favourable possibility would be 2+28 sec,
but {\bf any solution that considers one snapshot shorter than 15 sec would be an improvement}.
To make the analysis easier and faster, we would then need that {\bf photometry of the single snapshots be provided} in addition to that of the combined snapshots. This would avoid heavy user-generated processing. 
In any case, we stress the importance of developing methods to manage saturation, like the ``halo photometry" developed for Kepler/K2 images (White et al. 2017), which are expected to partly mitigate the problem.

While the study of blazar variability requires especially good sampling, the analysis of blazar environment needs depth. Indeed, to check the presence of a galaxy cluster in the blazar location we must verify the existence of a ``red sequence" in the colour versus magnitude diagram, with the right colour expected for the blazar redshift. The possibility to make several plots of this kind increases the robustness of the results.
To analyse the clustering properties up to redshift $\sim 2$ we must reach a magnitude of about 25 at 1.4 micron (Kotyla et al. 2016).
By considering an elliptical galaxy template, we estimate that this means 26.5 in the $z$ band, which is approximately what the WFD will reach in 10 years. However, the same analysis within redshift 1 requires a depth of $\sim 25$ in the $z$ band, and within redshift 0.5 a depth of 24. Therefore, the study of AGN environment with LSST will progressively cover more distant regions as time passes, starting from the local Universe with the first-year data, up to cosmological distances in the later stages of the project. And along the way we will have the opportunity to refine the analysis methods to tackle the increasing challenges that we expect when moving towards higher redshifts.

Taking into account the above points, we consider as highly desirable a {\bf rolling cadence} like the ones proposed in the pontus\textunderscore 2579, rolling\textunderscore 10yrs or rolling\textunderscore mix\textunderscore 10yrs OpSim Runs, including 2 alternating declination bands with increased observations (75\% of the WFT time) of half the sky for some time, at the expenses of the other half, which would have increased sampling for the rest of the time. The above OpSim Runs consider alternating over- and under-sampling every year. However, in this way some sources would never have a complete observing season with the same sampling. We notice that {\bf a rolling cadence with a turnover of at least 18 months} would overcome this inconvenience.
The rolling cadence would allow us to obtain better sampled light curves, even if in alternate periods only, to follow blazar activity and discover new blazars on the base of their variability. 

The observing strategies described above could also fit the FRBs case, if the possible optical counterparts were transients with time scales similar to the blazar variability time scales.
It could also fit the unidentified $\gamma$-ray sources, at least those that will reveal themselves to be blazars.

{\bf Other cadences}

At this stage, we cannot make a case for a multimessenger synergy with neutrino telescopes, since detection of high-energy neutrinos of astrophysical nature is still a rare circumstance and because we currently believe that the most likely neutrino blazar sources are bright objects (e.g. Righi et al. 2018). However, the scenario is still unclear (the uncertainty in the neutrino direction can be up to a few degrees), and in the future we might need to explore other possibilities that only LSST would definitely ascertain. If this will be the case, we would like to have the opportunity to ask for {\bf ToO observations} triggered by a neutrino alert.

We finally consider as very interesting the {\bf star trail cadence} proposed in the white paper by David Thomas et al. Indeed, even if we do not expect flux changes of the order of a mag on subsecond time scales for blazars, the star trail method would remove saturation and thus allow us to follow both sources during outburst phases and sources that are usually bright, as the present blazar neutrino sources candidates. 

We mention that the star trail cadence would be of extreme interest for FRBs, in case the optical counterpart had a time duration similar to the radio event (tens of milliseconds) and intrinsic optical luminosity of the order of ten times that of the 1--3 Gpc distant host galaxy (a strong requirement, but still in line with some FRB models). In this case LSST would see some of these events, which would be an outstanding result.
 
For the above reasons, we propose a {\bf minisurvey cadence linking bright/flaring blazars with the search of extremely fast (subsecond) transients through a star trail technique}. 
We can roughly estimate that of the known ~5000 blazars, about 2000 will be seen by LSST in the southern sky and about 10\% of them will be brighter than the saturation limit for a 15 sec snapshot\footnote{We are considering here the $r$-band case for which we have more information.} (see the bottom-right figure in Section 2). This means that about 200 blazars will saturate either because they are always very bright or because they are observed in a flaring state. By considering an optical observing season of about 6 months, every WFD complete map of the sky will contain $\sim 100$ saturated blazars. Assuming that a complete map is done every about 5 days, and taking into account that blazars are uniformly distributed in the sky (but in the Galactic Plane, where we cannot see them), this means $\sim 20$ saturated blazars a day in different fields. If saturation could be recognized ``on-the-fly" by the LSST prompt pipeline during the first visit of a field, then we could ask that, if saturation affects selected targets (here: blazars), then the second visit of the same field during the same night be performed with a modified observing mode. This could simply be, as proposed above, a visit with 2+28 sec exposure, but, more interestingly, it could be a ``star trail" mode with tracking turned off. The star trail image could possibly be taken in addition to the normal planned second visit. If the exposure time of a star trail image were 15 sec, then every night LSST would dedicate 5 minutes (net exposure) to observe $\sim 20$ fields with the star trail mode.
In one year, LSST would have spent about 30 hours to get some 7000 star trail images distributed over more than 200 fields (the flaring state can shift from one blazar to another) to look for extremely fast transients, with the by-product of helping blazar science. The result would be a $\geq 2000$ square degree star trail minisurvey, large enough and deep enough to maximize the possibility to discover subsecond transients. No other smaller telescope could do the same.


\subsection{High-level description}
WFD: Rolling cadence with 2 alternating declination bands, with 75\% of WFD time on half sky and 25\% on the other half, possibly with a 18 month turnover.
2 or 3 visits per night, one of which with a reference filter (e.g. $r$), 2 snapshots/visit with diversified exposures (e.g. 2+28 sec).

Minisurvey: star trail exposure in the second visit of the same night for the fields where a saturated blazar is detected in the first visit

\vspace{.3in}

\subsection{Footprint -- pointings, regions and/or constraints}
Star trail observing mode must avoid crowded fields.

\subsection{Image quality}
No particular constraint on image quality.

\subsection{Individual image depth and/or sky brightness}
We would retain the total exposure time of 30 sec/visit to maintain the current depth/visit, but spread on two snapshots with diversified exposure times (e.g.: 2+28 sec).

\noindent
No particular constraint on the sky brightness.

\subsection{Co-added image depth and/or total number of visits}
The co-added image depth is most important for environmental studies.
Deeper means higher redshift. We would need to reach mag $\sim 25$ in the $z$ band to explore up to a redshift of about 1.

\subsection{Number of visits within a night}
At least 2 visits, one of which in a reference filter (e.g.\ $r$). 

\subsection{Distribution of visits over time}
The 2 (or 3) visits per night are in general more useful if they are spaced in time, but this is not so critical. A time interval of 3--4 days between subsequent observations of the same field would lead to a fair sampling of normal blazar activity, but not of flaring states. Also for this reason we favour a rolling cadence, which can give a better sampling, although only in certain time periods.

\noindent
Blazar monitoring is more efficient when {\bf prolonged as far as possible during the observing season} to reduce gaps in the light curves. This also favours simultaneous observations with satellites that can point at the source only at the boundaries of the optical observing season, like XMM-Newton. Therefore, we favour observations also at relatively high air masses.

\subsection{Filter choice}
As stressed before, we need a reference filter to follow in detail the source behaviour with reliable light curves, not affected by errors due to colour corrections or any kind of interpolation. This reference filter could be the $r$ one. 
All the other filters are important to define the blazar spectrum and hence the blazar type.

\subsection{Exposure constraints}
As explained above, saturation is a strong issue, as blazars can undergo outbursts involving several magnitude brightening. Moreover, some interesting sources that are high-energy photons and neutrino emitters and will be the targets of, e.g., CTA observations, are generally bright.
Single snapshots of 30 sec each would foil many blazar studies.
Twin snapshots of 15 sec each would still be very limiting.
Therefore, we propose to preserve the 2 snapshots/visit with a total 30 sec exposure to go deep, but to diversify the two exposures to reduce the saturation limit on the shortest exposure. This should {\bf possibly be below 6 sec}, an exposure that would lower the saturation limit to $r=15$.

\subsection{Other constraints}

\subsection{Estimated time requirement}
The proposed general strategy for the WFD does not require additional time, while 
the star trail minisurvey triggered by saturated blazars would require about 5 minutes/day, i.e.\ 30 hours/year.

\vspace{.3in}

\begin{table}[ht]
    \centering
    \begin{tabular}{l|l|l|l}
        \toprule
        Properties & Importance \hspace{.3in} \\
        \midrule
        Image quality &  3   \\
        Sky brightness &  3 \\
        Individual image depth & 3  \\
        Co-added image depth &  1 \\
        Number of exposures in a visit   &  1 \\
        Number of visits (in a night)  & 2  \\ 
        Total number of visits &  1 \\
        Time between visits (in a night) &  2 \\
        Time between visits (between nights)  & 1  \\
        Long-term gaps between visits & 1 \\
        Exposure time & 1 \\
        \bottomrule
    \end{tabular}
    \caption{{\bf Constraint Rankings:} Summary of the relative importance of various survey strategy constraints. Please rank the importance of each of these considerations, from 1=very important, 2=somewhat important, 3=not important. If a given constraint depends on other parameters in the table, but these other parameters are not important in themselves, please only mark the final constraint as important. For example, individual image depth depends on image quality, sky brightness, and number of exposures in a visit; if your science depends on the individual image depth but not directly on the other parameters, individual image depth would be `1' and the other parameters could be marked as `3', giving us the most flexibility when determining the composition of a visit, for example.}
        \label{tab:obs_constraints}
\end{table}

\subsection{Technical trades}

We favour a larger number of visits rather than a larger survey area.

Our science case would benefit from a rolling cadence.

We favour an increase of the overall number of visits. Indeed, an increase of the individual image depth, with the correlated increase of the saturation limit, would seriously hinder our science case.

We do not need uniformity in the number of visits in the different filters.

\section{Performance Evaluation}

We identified the following Metrics:


a) Inter-night gap - should ideally be not longer than 3 days in the reference filter, to have at least one light curve sampled enough to recognize blazars and to promptly detect flaring episodes and trigger follow-up observations; this could be accomplished even only in some years (rolling cadence).

We used MAF with TransientMetric on various databases (baseline2018a- the Project-official baseline, against pontus\textunderscore 2579, rolling\textunderscore 10yrs, rolling\textunderscore mix\textunderscore 10yrs, all including rolling cadences with 2 declination bands alternating every year and 75\% and 25\% over and undersampling) to check the performance of different cadences in revealing flaring events. We simulated cases with 0.5/1/2 mag increase and decrease in 1/2/4 weeks and investigated the following metrics:

b) number of detected flares; 

c) number of detected flares on rise (demanding 2 points before $\rm t_{max}$) to be able to trigger follow-up observations.

We found that the pontus\textunderscore 2579 gives the best results, as shown in the table below. 
The comparison between rolling\textunderscore 10yrs (considering the two visits of the same night with the same filter) and rolling\textunderscore mix\textunderscore 10yrs (with different filters) shows that the first is superior in fraction of events detected on rise, while the latter is better in fraction of alerts.

\vspace{0.3in}
\begin{tabular}{l l l l l l}     
\hline
OpSim Run     & $\Delta r$ & $\Delta t$ & $r_{\rm min}$ & Alert (\%) & Detected on rise (\%)\\ 
\hline
baseline\textunderscore 2018a & 0.5        & 7          & 18            & 23.8 & 12.3  \\
pontus\textunderscore 2579    & 0.5        & 7          & 18            & {\bf 28.8} & {\bf 12.7}   \\
rolling\textunderscore 10yrs  & 0.5        & 7          & 18            & 23.5 & {\bf 12.7}   \\
rolling\textunderscore mix\textunderscore 10yrs & 0.5      & 7          & 18            & {\bf 28.8} & 10.8   \\
\hline
baseline\textunderscore 2018a & 1        & 14          & 18            & 36.9 & 21.5   \\
pontus\textunderscore 2579    & 1        & 14          & 18            & {\bf 46.9} & {\bf 23.1}   \\
rolling\textunderscore 10yrs  & 1        & 14          & 18            & 35.4 & 20.0   \\
rolling\textunderscore mix\textunderscore 10yrs & 1      & 14          & 18            & 38.5 & 10.8   \\
\hline
baseline\textunderscore 2018a & 2        & 30          & 18            & 48.3 & 35.0   \\
pontus\textunderscore 2579    & 2        & 30          & 18            & {\bf 63.3} & {\bf 41.7}   \\
rolling\textunderscore 10yrs  & 2        & 30          & 18            & 48.3 & 35.0   \\
rolling\textunderscore mix\textunderscore 10yrs & 2      & 30          & 18            & 50 & 33   \\
\hline
\end{tabular} 

\vspace{0.3in}
The figure below shows the results obtained with the pontus\textunderscore 2579 OpSim Run in the first case examined, that of a 0.5 mag flare with a time scale of 1 week.

\begin{center}
\includegraphics[scale=0.5]{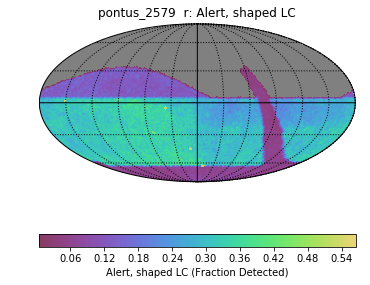}
\includegraphics[scale=0.5]{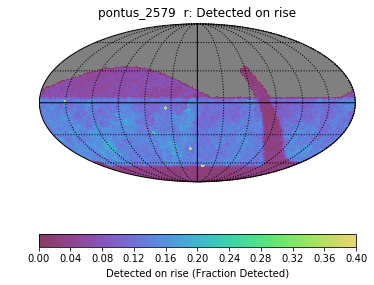}
\end{center}

d) Max airmass - to reduce gaps in the light curves, we need to prolong the observing season as far as possible, i.e. to observe also at high air masses
Also here the pontus\textunderscore 2579 OpSim Run seems better than baseline\textunderscore 2018a (2.75 against 1.51 for the WFD all bands).

e) Final depth - for environmental studies the deeper we can go, the farther in redshift we can explore. The fiveSigmaDepth Metrics in baseline\textunderscore 2018a and pontus\textunderscore 2579 are similar.

\vspace{.6in}

\section{Special Data Processing}

{\bf Photometry of single snapshots}. If LSST is not going to provide photometry of single snapshots, as it seems, we will need to run pipelines on image cutouts to perform it.

{\bf Photometry of saturated sources}. Saturation is a serious issue for studies of blazar variability with LSST. We have proposed methods that can avoid saturation either by adopting different snapshot exposures or by turning tracking off. Both of them can be interesting also for the study of short-time transients. However, if this cannot be achieved, we will try to develop methods to approximate the real flux of saturated sources using the wings of the point spread function. This will give at least a rough estimate of the source brightness.

{\bf Photometry of sources in star trail images}. The star trail method is proposed by David Thomas et al. in another white paper. They have already developed specific methods to analyse these images. Additional effort will be devoted to obtain accurate photometry for the trails.

\section{Acknowledgements}
This work was developed within the Transients and Variable Stars Science Collaboration (TVS) and the authors acknowledge the support of TVS in the preparation of this paper.
We thank Federica Bianco and Rachel Street for their very valuable work as TVS chairs and David Thomas, Melissa Graham, Gordon Richards as well as many others who participated on the Slack and Community forums for useful discussions. We acknowledge financial support from our Istituto Nazionale di Astrofisica (INAF, Italy) and from the LSST Corporation (LSSTC) - LSST Enabling Science Grants.

\section{References}

Dey et al. 2018, ApJ, 866, 11

\noindent
Ghisellini et al. 2013, MNRAS, 432, 2818

\noindent
Keane et al. 2018, MNRAS, 473, 116

\noindent
Kotyla et al. 2016, ApJ, 826, 46

\noindent
Magliocchetti et al. 2018, MNRAS, 478, 3848

\noindent
Raiteri et al. 2012, A\&A, 545, A48

\noindent
Raiteri et al. 2017, Nature 552, 374

\noindent
Righi et al. 2018, MNRAS, in press, arXiv:1807.04299

\noindent
Spitler et al. 2016, Nature, 531, 202

\noindent
The IceCube Collaboration et al. 2018, Science, 361, 6398

\noindent
White et al. 2017, MNRAS, 471, 2882

\end{document}